\documentstyle[twoside,fleqn,espcrc2,epsfig,amsfonts,latexsym]{article}

\newcommand{\AmS}{{\protect\the\textfont2
  A\kern-.1667em\lower.5ex\hbox{M}\kern-.125emS}}
  \newcommand{\bbox}{\lower0.85pt\hbox{$\Box$}}
  \newcommand{\dual}{\mbox{}^{\ast}}
  \newcommand{\kreisl}{\raise0.85pt\hbox{$\scriptstyle\bigcirc$}}
  \newcommand{\dreieck}{\raise0.85pt\hbox{$\scriptstyle\bigtriangledown$}}
  \newcommand{\stern}{\lower0.85pt\hbox{\Large $\star$}}

\title{
\vspace{-8mm}
\rightline{\small UL-NTZ 22/98, KANAZAWA-98-13, ITEP-TH-44/98}
\vspace{-2mm}
\rightline{\small September 5, 1998}
	Embedded Topological Defects in Electroweak Theory:\\ 
        From Percolating Networks to Sphalerons\thanks{
	Combined contribution covering the talks at LAT'98 by M.~N.~Chernodub 
	and E.-M.~Ilgenfritz}
      }

\author{
	M.~N.~Chernodub\address{ITEP, B. Cheremushkinskaya 25, 
	Moscow 117259, Russia}, 
	F.~V.~Gubarev$\mathrm{^a}$,
	E.-M.~Ilgenfritz\address{Institute for Theoretical Physics,
	Kanazawa University, Kanazawa 920-1192, Japan}
	and A.~Schiller\address{Institut f\"ur Theoretische Physik and NTZ,
	University of Leipzig, D-04109 Leipzig, Germany}
       }
       
\begin{document}

\begin{abstract}
New defects ($Z$-vortices and Nambu monopoles) are found to become 
thermodynamically relevant for the broken phase near to the 
(weakly first order) electroweak phase transition, and below the
crossover for higher Higgs mass. 
The symmetric phase is characterized by vortex condensation (percolation). 
The percolation transition persists in the crossover region.
The quasiclassical nature of the vortices is supported by correlations with 
Higgs field and gauge field energy. Sphalerons are shown to be 
related to monopolium bound states.
\end{abstract}

\maketitle

\section{INTRODUCTION}
Is the electroweak phase transition accompanied by the condensation
of some defects? 
What is happening with the defects at the crossover region?
Is there a range of parameters supporting a 
dilute gas of such defects in the broken phase and 
how are they related to the well-known sphaleron configurations? These questions 
have been explored during the last year within the $SU(2)$ Higgs model. 

The $Z$-vortex string~\cite{Ma83,Na77} corresponds to the 
Abri\-ko\-sov--Niel\-sen--Ole\-sen~\cite{ANO} vortex solution related to
the Abelian subgroup of $SU(2)$ embedded into the electroweak
theory~\cite{VaBa69,BaVaBu94}. $Z$-strings have been demonstrated to be 
unstable
both at zero and at finite temperatures~\cite{HolmanVa}.
Sphalerons~\cite{Ma83,DaHaNe74,KlMa84} are unstable
configurations as well, being transition states (saddle 
points)
between vacua with different Chern-Simons number.
Classical sphalerons have been constructed on the lattice in 
Refs.~\cite{vanBaal} by a specific saddle point cooling technique
(these configurations have been important for our project).
Baryon asymmetry should 
not be washed out by sphaleron processes after the electroweak symmetry 
breaking has been completed. This 
requirement
puts stringent conditions 
on the strength of the phase transition because the activation energy 
of these barriers strongly depends on the strength of the electroweak 
phase transition~\cite{SphalRev}.

Recently, lattice studies have shown (see ~\cite{laineLatt98} for a
conclusive summary) that neither the simplified $SU(2)$ Higgs model nor the 
$SU(2) \times U(1)$ model have a thermal transition
of the necessary strength, given the experimental bounds on the Higgs mass.
This observation has triggered our interest to investigate whether,
for Higgs masses below the end of the phase transition~\cite{wePRD} and in the 
crossover region above, topological excitations might be excited with a 
non-negligible statistical weight. These historical circumstances 
might explain why defects in the context of the electroweak phase transition 
are a rather new topic within lattice gauge theory.

In this contribution we briefly describe our lattice definitions and 
present selected results, some of them having not been published before. 
Details can be found in Refs.~\cite{weSphal,defect70,defect100}. 
Our study has been performed within the dimensionally reduced $SU(2)$ Higgs 
theory which provides an effective representation of the electroweak theory 
in some range of Higgs masses and at temperatures near to the electroweak scale. 
For the formulation of the lattice version of the effective theory see the
talk of A. Schiller at this conference~\cite{SpectrumLat98}. 
The lattice gauge coupling $\beta_G=4/(a~g_3^2)$ (with $g_3^2 \approx g_4^2~T$)
gives the lattice spacing in units of temperature, and the hopping parameter 
$\beta_H$ substitutes $m_3^2$ driving the transition. The parameter of the 
phase transition or the crossover can be translated into temperature $T_c$ 
and a Higgs mass $M_H\approx M_H^*$ (see {\it e.g.} Ref.~\cite{weNPB} and
references therein). Notice that the Higgs field is 
usually written as a multiple of a $SU(2)$ matrix $\Phi$, here we need it 
in the form of a $2$-component isospinor~$\phi$.

\section{DEFINITIONS}

The gauge invariant and quantized lattice definition~\cite{weSphal} of 
elementary $Z$-vortices and Nambu monopoles is closely related to the 
definition in the continuum theory~\cite{Na77}.
First we define a composite adjoint unit vector field
\begin{equation}
n_x = n^a_x \sigma^a \,,\quad
n^a_x = - \frac{\phi^+_x \sigma^a \phi_x}{\phi^+_x \phi_x}\,,
\nonumber
\end{equation}
$\sigma^a $ are Pauli matrices. For the following purpose, $n_x$ 
plays a role similar to the adjoint Higgs field in the
definition of the 't~Hooft--Polyakov monopole~\cite{tHPo74} within the
Georgi--Glashow model. It helps to define the gauge invariant 
flux ${\bar \theta}_p$ through the pla\-quet\-te $p=\{x,\mu\nu\}$,
\begin{equation}
{\bar \theta}_p =  \arg \Bigl( {\mathrm {Tr}}
\left[(1 + n_x) V_{x,\mu} V_{x +\hat\mu,\nu}
V^+_{x + \hat\nu,\mu} V^+_{x,\nu} \right]\Bigr)
\label{AP}
\end{equation}
via the projected links
\begin{equation}
V_{x,\mu}(U,n) =  U_{x,\mu} + n_x U_{x,\mu} n_{x + \hat\mu}\,.
\nonumber
\end{equation}
The auxiliary Higgs field $n_x$ is covariantly constant with respect to 
these $SU(2)$-valued links.
Originally, Nambu monopoles are topological defects of an Abelian field
which is defined in the unitary gauge ($\phi_x={(0,\varphi)}^T$ and 
$n^a_x \equiv \delta^{a3}$) by the phases $\theta^u_l = \arg U^{11}_l$. 
This field behaves as a compact Abelian field with respect to residual 
Abelian gauge transformations which leave the unitary gauge condition intact.
The gauge invariant flux would be extracted from its pla\-quet\-tes $\theta^u_p$
as usual: 
${\bar \theta}_p = \left( \theta^u_p - 2 \pi m_p \right) \in [-\pi,\pi)$.
The magnetic charge of Nambu monopole cubes and the $Z$-vorticity number
of pla\-quet\-tes can alternatively be defined in a gauge invariant 
way~\cite{weSphal} referring to (\ref{AP}). The monopole charge $j_c$ 
inside
the cube $c$ is defined in terms of the gauge invariant fluxes 
(\ref{AP}) penetrating the surface $\partial c$:
\begin{equation}
j_c = - \frac{1}{2\pi} \sum_{p \in \partial c}
{\bar \theta}_p\,. 
\label{jN}
\end{equation}
The $Z$--vorticity number of the pla\-quet\-te $p$
is defined~\cite{weSphal} as follows:
\begin{equation}
\sigma_p = \frac{1}{2\pi} \Bigl( \chi_p - {\bar \theta}_p \Bigr) \,,
\label{SigmaN}
\end{equation}
where ${\bar \theta}_p$ has been given in (\ref{AP}), and
\begin{equation}
\chi_{p} = \chi_{x,\mu\nu} = \chi_{x,\mu} + \chi_{x +\hat\mu,\nu} -
\chi_{x + \hat\nu,\mu} - \chi_{x,\nu}\,,
\nonumber
\end{equation}
is a pla\-quet\-te variable written in terms of the Abelian links
$\chi_{x,\mu} = \arg\left(\phi^+_x V_{x,\mu} \phi_{x + \hat\mu}\right)$.

A $Z$--vortex is formed by links $l=\{x,\rho\}$ of the dual
lattice which are dual to pla\-quet\-tes $p=\{x,\mu\nu\}$ which carry
a non-zero vortex number (\ref{SigmaN}): $\dual \sigma_{x,\rho} =
\varepsilon_{\rho\mu\nu} \sigma_{x,\mu\nu} \slash 2$. 
$Z$--vortex trajectories
are either closed or begin/end on Nambu (anti-) monopoles:
\begin{equation}
\sum^3_{\mu=1} (\dual \sigma_{x-\hat\mu,\mu} - \dual \sigma_{x,\mu})
= \dual j_x\,.
\end{equation}
This is how we actually localize Nambu monopoles. 

When the embedded defects are physical excitations with some intrinsic 
size, one should be able to describe them on finer lattices as
{\it extended} topological objects.   
An extended monopole (vortex) of physical 
size $k~a$ is defined on $k^3$ cubes ($k^2$ pla\-quet\-tes, 
respectively)~\cite{IvPoPo90,Laine98}.
The charge of monopoles $j_{c(k)}$ on bigger $k^3$ cubes $c(k)$ is constructed
analogously to that of the elementary monopole (\ref{jN}) with
elementary $1\times 1$ plaq\-uet\-tes in terms of $V_{x,\mu}$
replaced by $n \times n$ Wilson loops (extended pla\-quet\-tes). In pure gauge 
theory, within the maximally Abelian gauge, 
these objects
are called 
type-I extended Abelian monopoles 
We have numerically investigated the behavior of the type-I extended
vortices of some {\it fixed} physical size on a series of finer and finer
lattices towards the continuum limit. Our results suggest that the
continuum limit for type-I extended vortices may exist. We remark that
an alternative, type-II, construction~\cite{IvPoPo90} of extended objects
fails to show a physically sensible behavior towards continuum limits.

\section{RESULTS}

We have scanned the phase transition with elementary 
defects at $M_H^*=30$ GeV (strong first order) and 70 GeV 
(weak first order) for $\beta_G=12$ 
on a relatively small lattice $16^3$.
At the lower Higgs mass we observed a discontinuity of the densities
$\rho_m = \sum\limits_c |j_c|/L^3$ of Nambu monopoles and 
$\rho_v = \sum\limits_p |\sigma_p|/(3~L^3)$ of vortices.
Simultaneously we have measured the percolation probability 
$C = \lim_{r \to \infty} C(r)$ derived from the cluster
correlation function~\cite{PoPoYu91}
\begin{equation}
C(r) = \frac{\sum\limits_{x,y,i}
\delta_{x \in \dual \sigma^{(i)}} \,\delta_{y \in \dual \sigma^{(i)}}
\cdot \delta\Bigl(|x-y|-r\Bigr)} 
{\sum\limits_{x,y} \delta\Bigl(|x-y|-r\Bigr)}\, ,
\end{equation}
where only those
connected clusters $\dual \sigma^{(i)}$ 
of vortex lines ($i$ labels distinct vortex clusters) contribute to the
nominator which pass
through the points $x$ and $y$.
Connected clusters $\dual \sigma^{(i)}$ are called percolating clusters 
if they contribute to the limit $C$. 
\begin{figure}[!htb]
\begin{minipage}{7.5cm}
\begin{center}
\epsfig{file=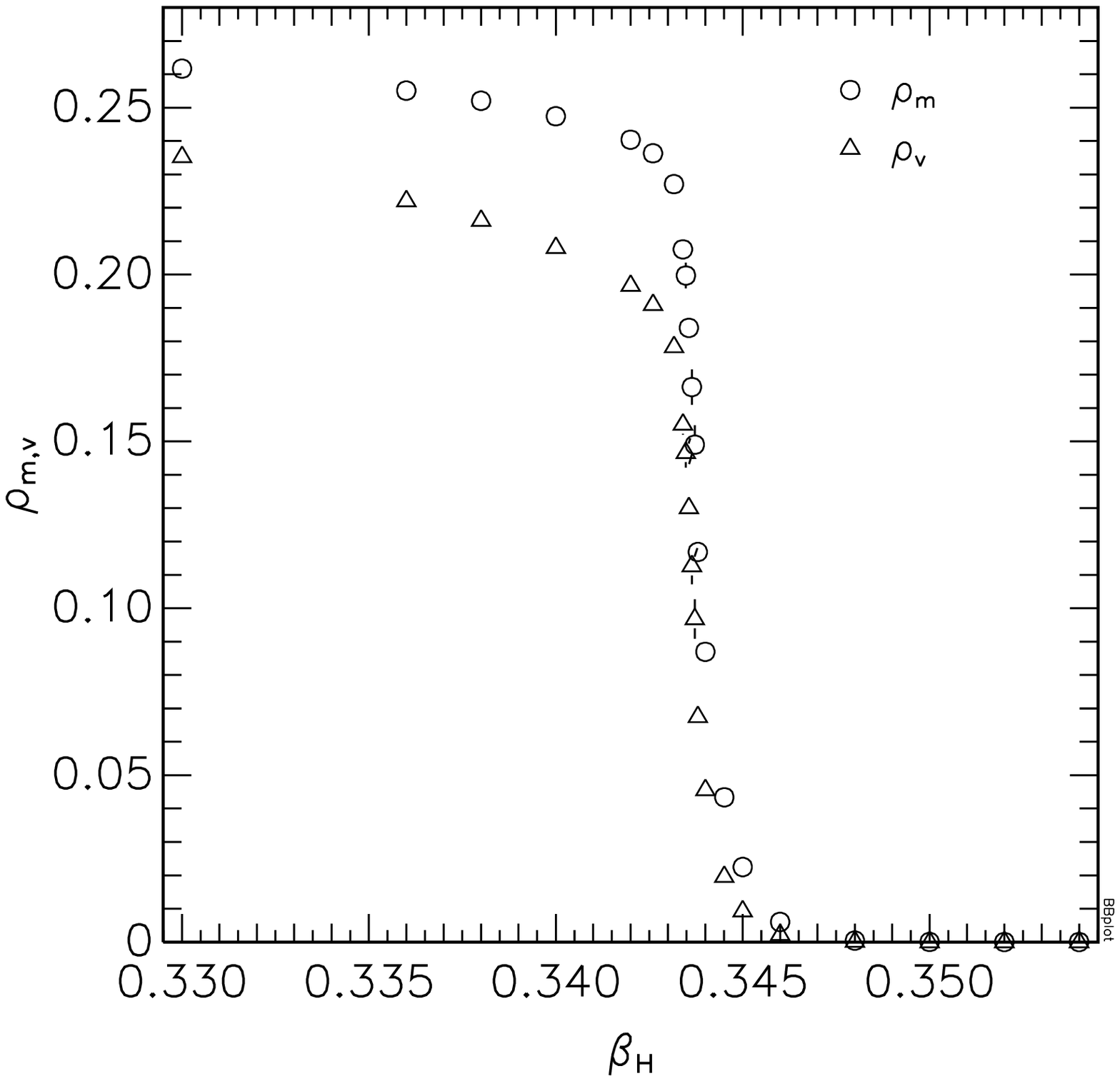,width=3.7cm,height=3.7cm,angle=0}
\epsfig{file=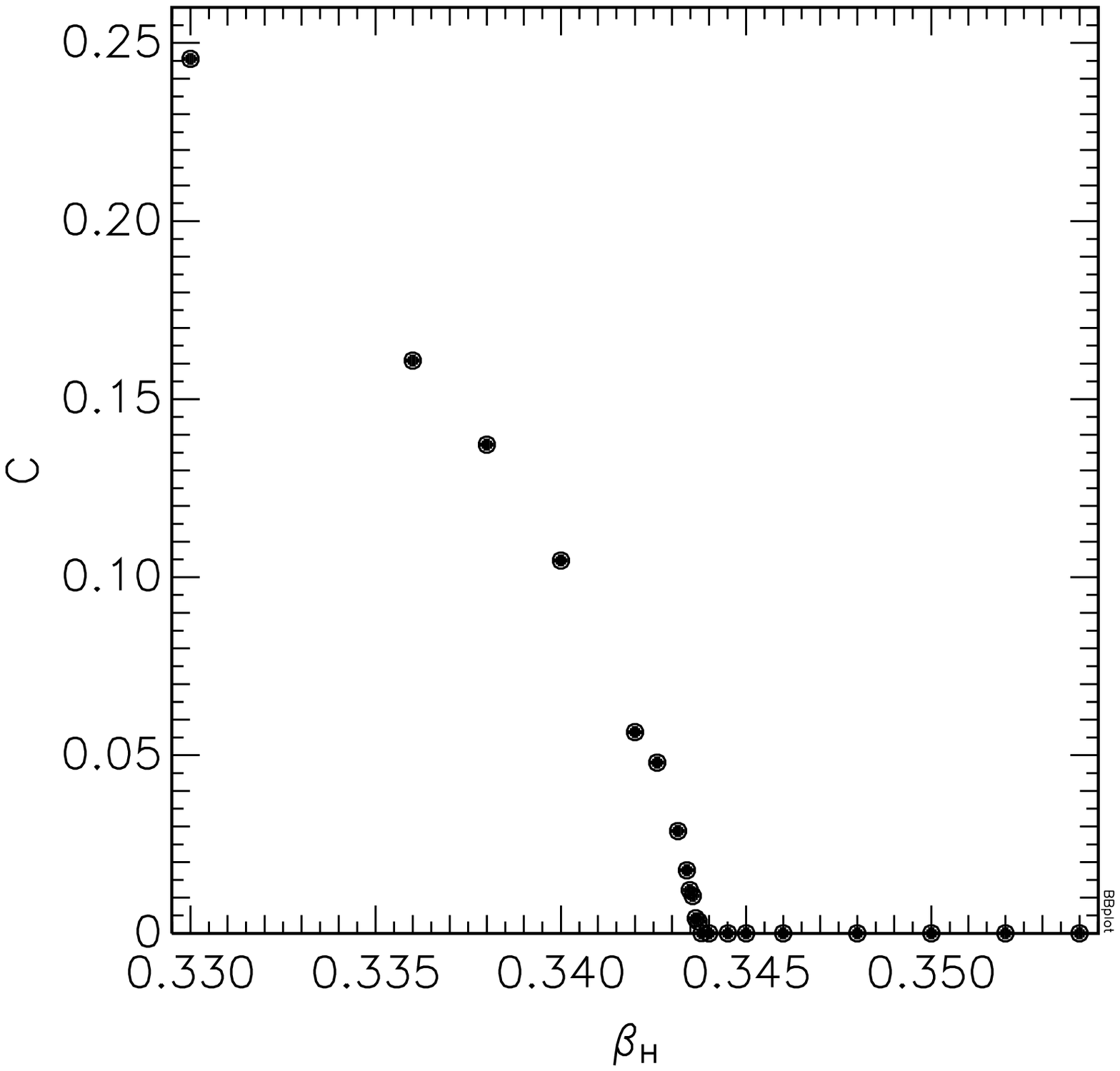,width=3.6cm,height=3.7cm,angle=0}
\vspace{-10mm}
\caption[]{\small 
Densities of elementary Nambu monopoles $\rho_m$ and
of $Z$--vortices $\rho_v$ (left)
for Higgs mass $M_H^*= 70$~GeV 
at gauge coupling $\beta_G = 12$; 
Percolation probability $C$ of
$Z$--vortex trajectories (right).}
\end{center}
\end{minipage}
\end{figure}

The percolation probability has a 
finite jump to zero at the pseudocritical $\beta_{Hc}=0.3411$ 
for $M_H^*=30$ GeV.  
The same study at $M_H^*=70$ GeV, already near to the 
endpoint of the first order transition, is summarized in Figure~1 and 
shows the percolation probability continuously vanishing towards 
$\beta_{Hc}=0.34355$. This is an inflection point of the densities $\rho_m$ and 
$\rho_v$ where they are approximately equal to half of their 
values in the symmetric phase. 
For $\beta_{H} > \beta_{Hc}$ the densities decrease exponentially. 

For the case $M_H^*=30$ GeV (where we find zero 
densities of elementary defects on the broken side) we have made some 
tests whether the densities of extended defects on the 
symmetric side admit a continuum limit. We have chosen elementary vortices 
and monopoles on a lattice simulated at 
$\beta^{(1)}_G=8$ and compared them with 
extended defects of size $k=2,3$ for simulations at 
$\beta^{(2)}_G=16$ and $\beta^{(3)}_G=24$,
respectively. A lattice size $16^3$ corresponding to a shrinking volume 
has been taken for all three cases because we did not expect severe finite 
size effects at such a strong transition. 
\begin{figure}[!htb]
\begin{minipage}{7.5cm}
\begin{center}
\epsfig{file=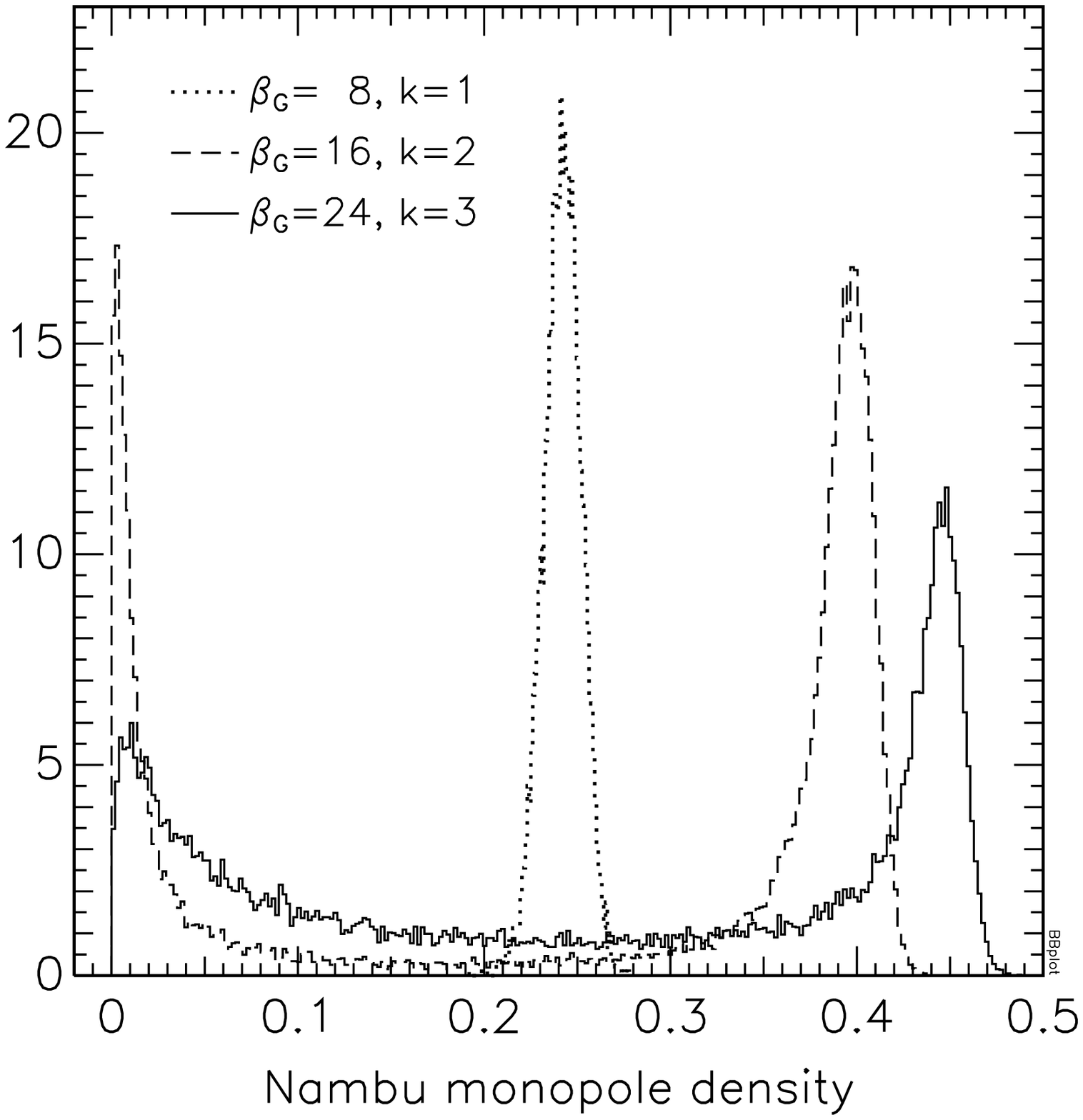,width=3.7cm,height=3.7cm,angle=0}
\epsfig{file=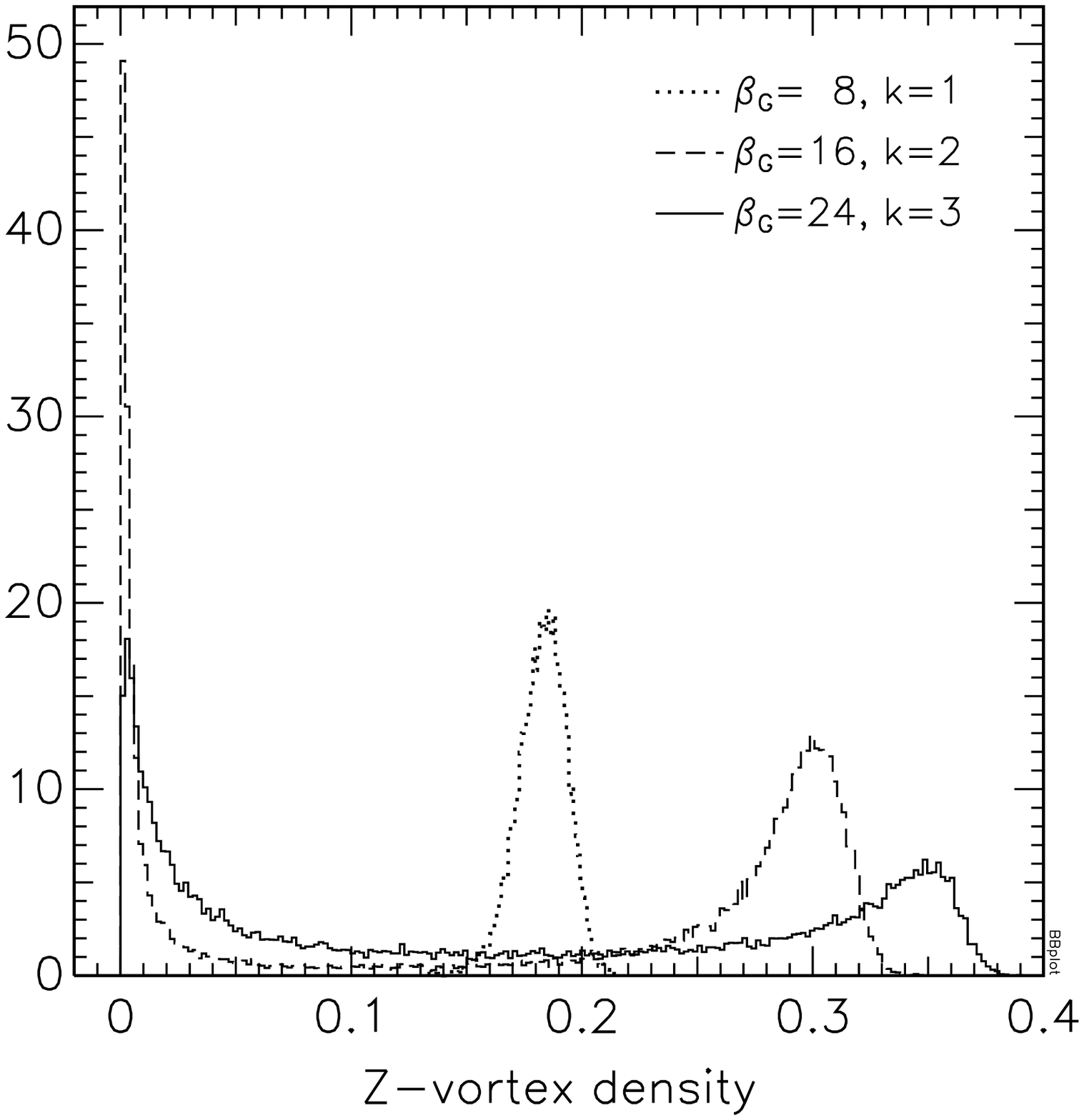,width=3.6cm,height=3.7cm,angle=0}
\vspace{-10mm}
\caption[]{\small
Density distributions of Nambu monopoles (left) and
$Z$--vortices (right) of fixed thickness at pseudocriticality
for different gauge couplings for Higgs mass $M_H^*=30$~GeV.}
\end{center}
\end{minipage}
\end{figure}

This (preliminary) test is shown in 
Figure~2 in the form of histograms of average vortex and monopole 
densities per configuration. 
The measurements have been performed near to the respective $\beta_{Hc}$ 
for the three values of $\beta_G$ and gave always a two-state signal with 
a density distribution peaking at or near to zero on the symmetric side. 
There is no way to perform an additive renormalization of the density at
the broken side by taking differences between the phases (in analogy to the
Higgs modulus squared $\rho^2$, 
Ref.~\cite{Laine98}).
With a box size $a(\beta_G=8)$ ($a\simeq 1/T$) we seem to be at the 
upper edge of the size distribution of physical vortices.
With increasing $\beta_G$, our construction for extended defects
seems to describe the vorticity better. This might explain the monotonous
increase to some (still unknown) {\it continuum} density of vortices/monopoles.

\begin{figure}[!htb]
\vspace{4mm}
\begin{minipage}{7.5cm}
\begin{center}
\epsfig{file=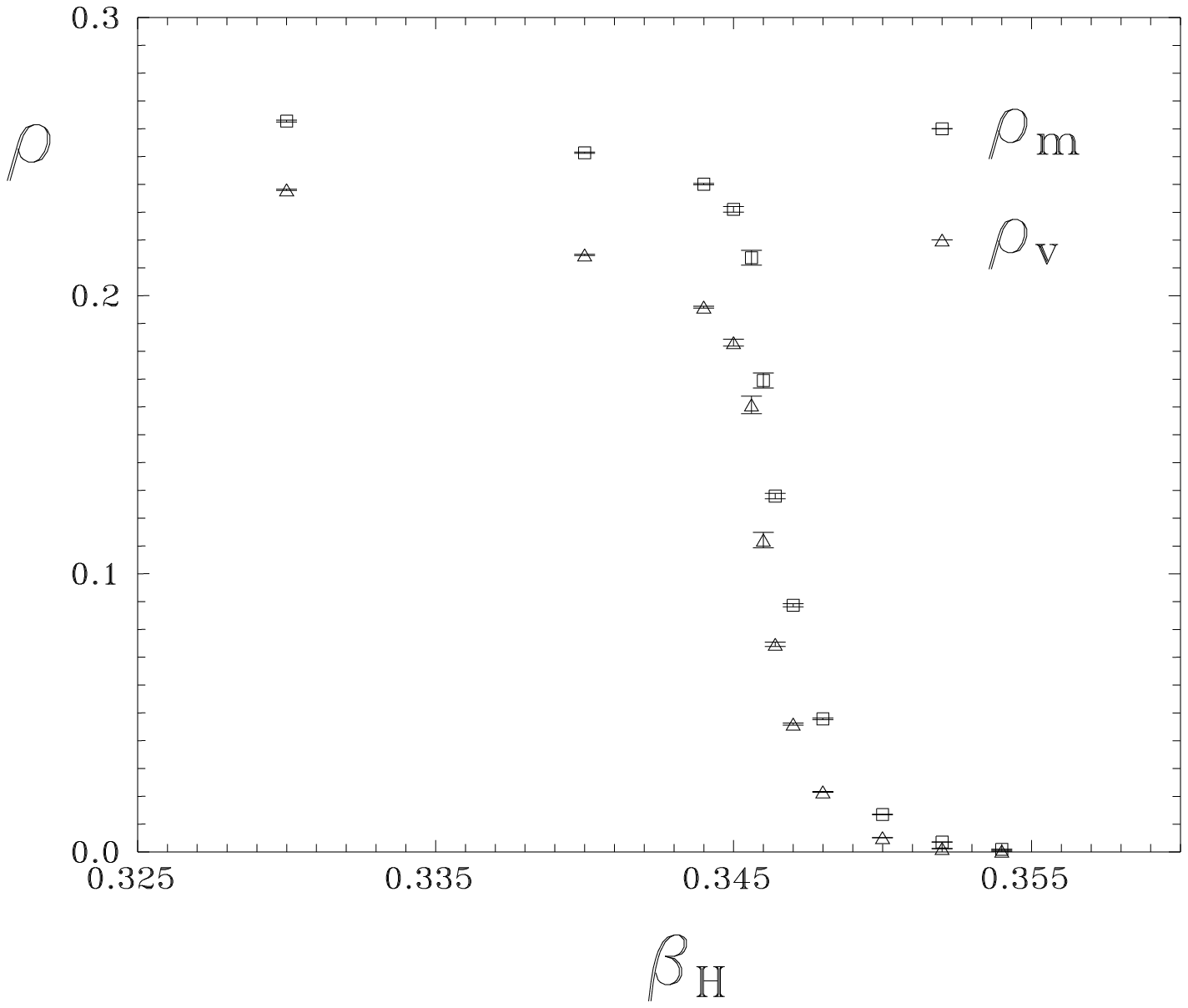,width=3.7cm,height=3.7cm,angle=0}
\epsfig{file=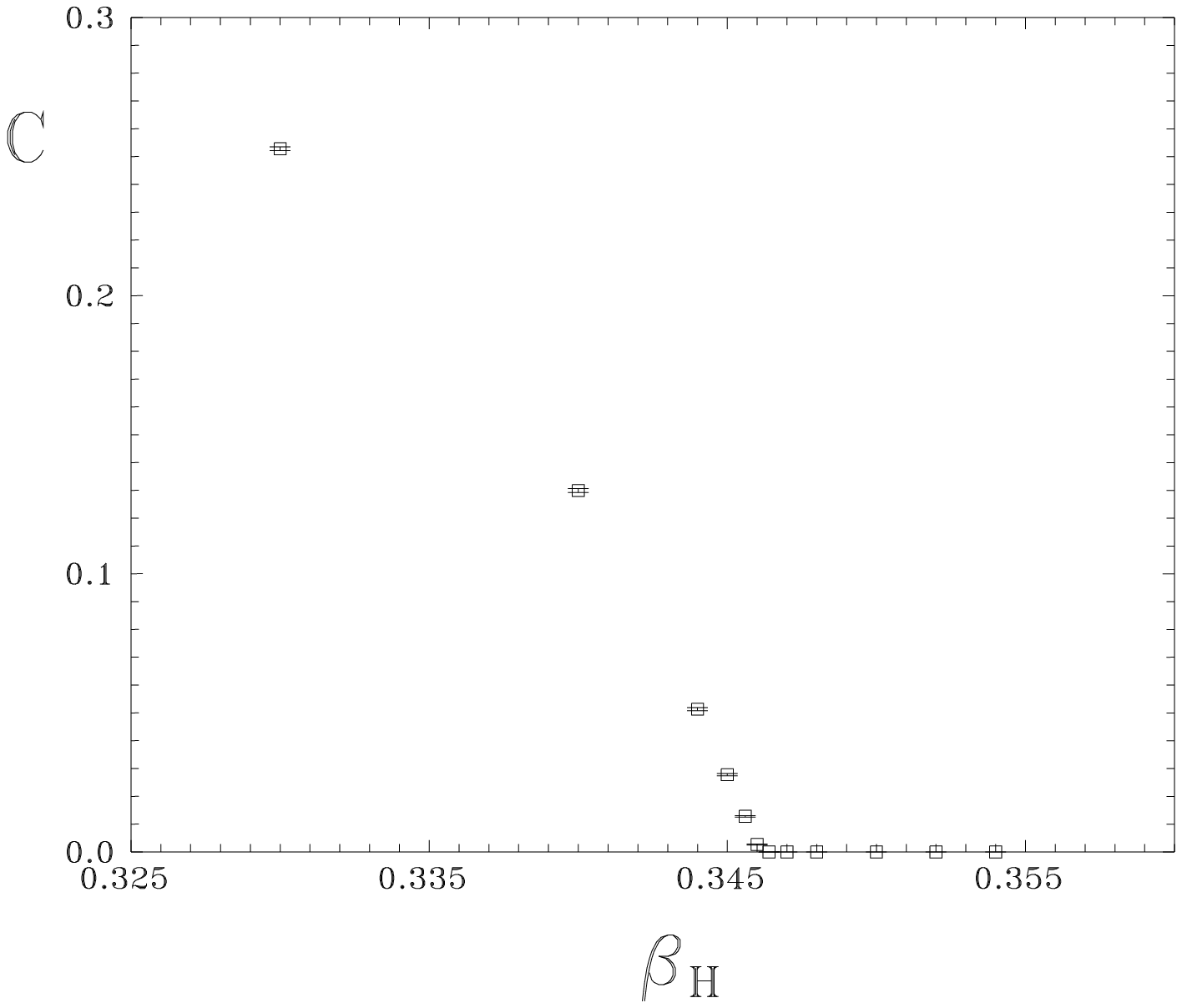,width=3.6cm,height=3.7cm,angle=0}
\vspace{-10mm}
\caption[]{\small 
Densities of elementary Nambu monopoles $\rho_m$ and
of $Z$--vortices $\rho_v$ (left)
for Higgs mass $M_H^*=100$~GeV 
at gauge coupling $\beta_G = 12$; 
Percolation probability $C$ of
$Z$--vortex trajectories (right).}
\end{center}
\end{minipage}
\end{figure}
\vspace{-10mm}
\begin{figure}[!htb]
\begin{minipage}{7.5cm}
\begin{center}
\epsfig{file=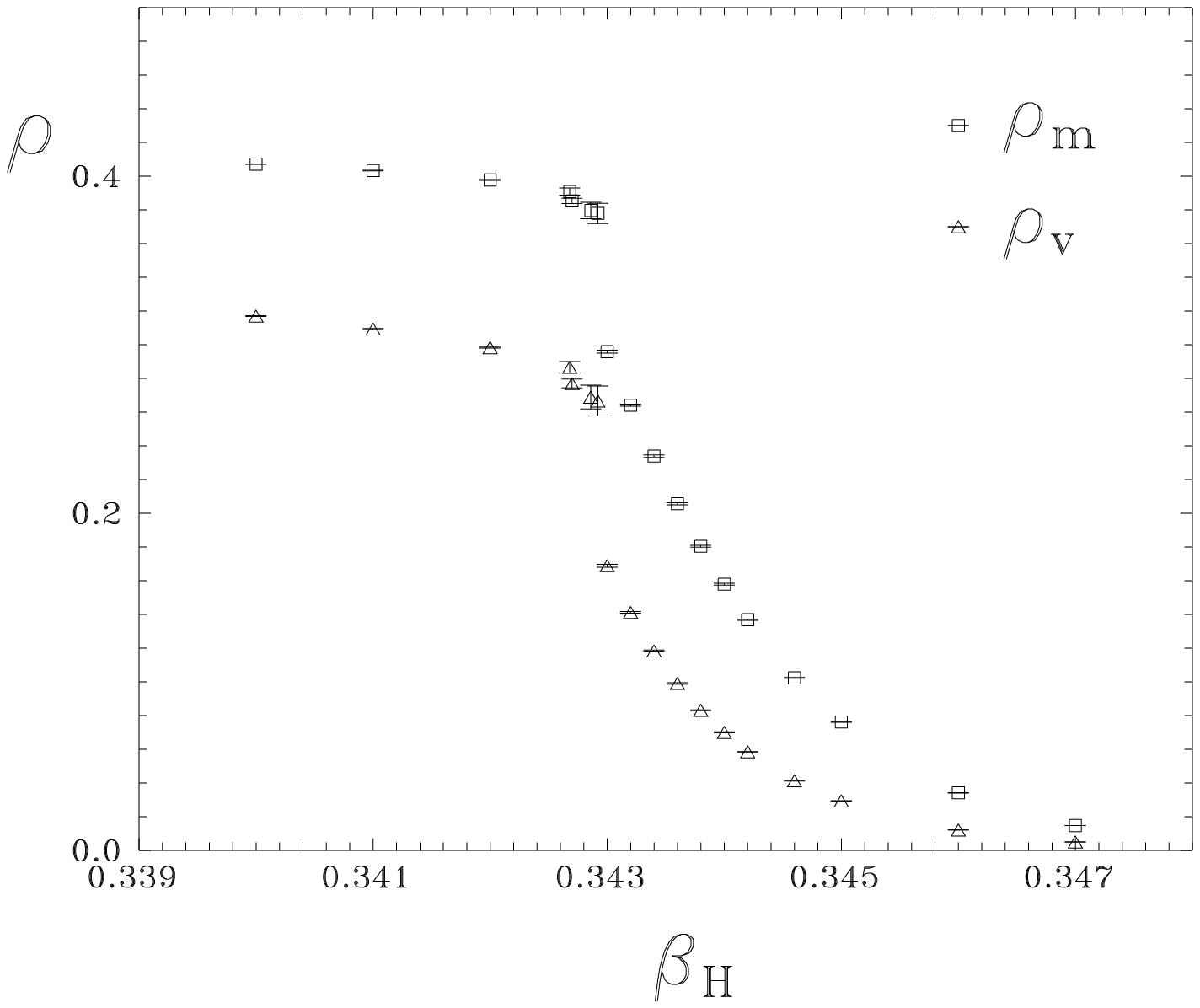,width=3.7cm,height=3.7cm,angle=0}
\epsfig{file=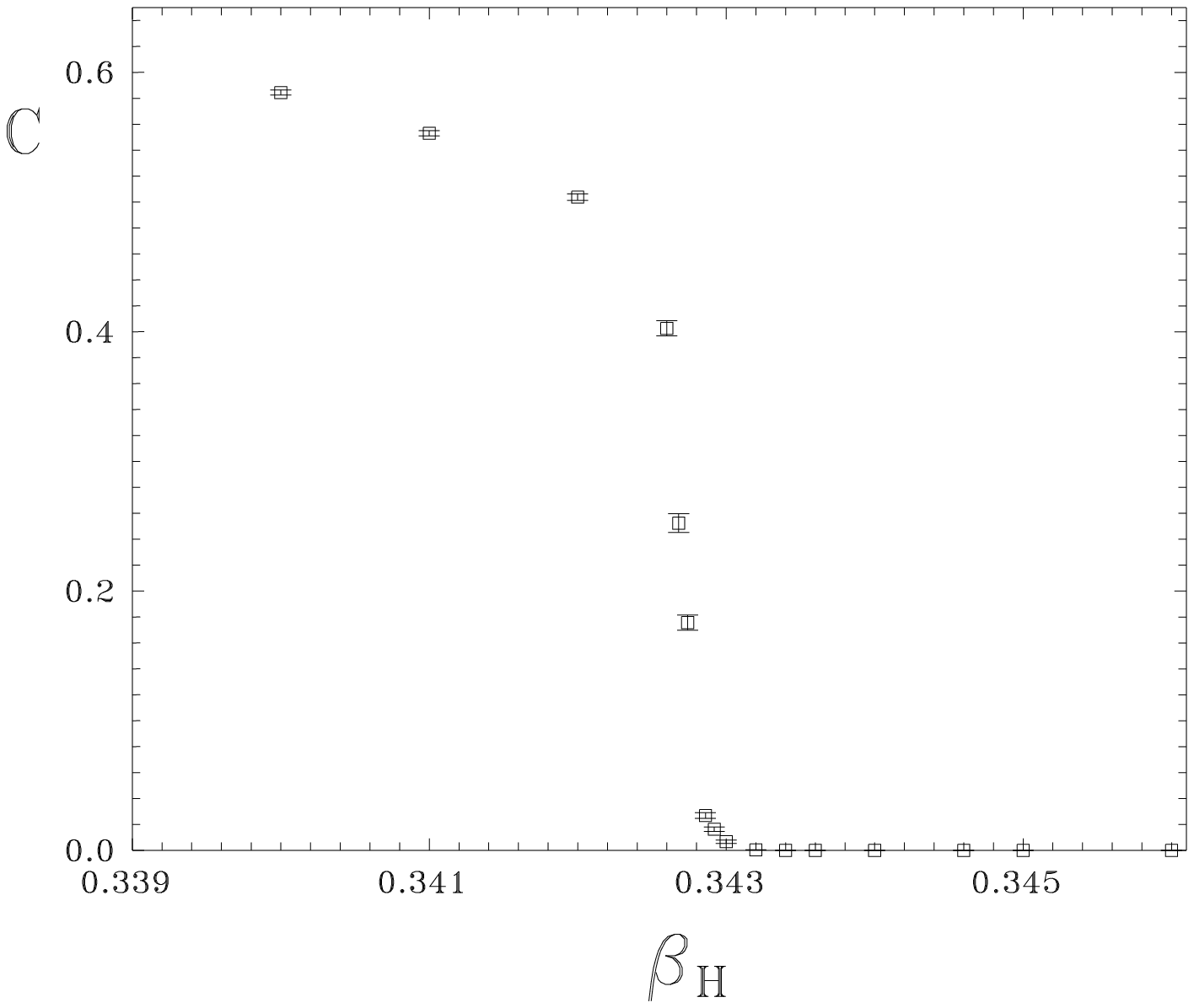,width=3.6cm,height=3.7cm,angle=0}
\vspace{-10mm}
\caption[]{\small
Same as Figure~3 for extended defects of 
size
$k=2$.} 
\end{center}
\end{minipage}
\end{figure}

Physically the crossover region is more interesting since we can 
study the properties of isolated defects in a not-so-dilute environment.
We show in Figure~3 the behavior of densities and percolation probability
$C$ {\it vs.} $\beta_H$ for $M_H^*=100$ GeV at $\beta_G=12$ for elementary
defects. 
We were wondering whether there is an universal percolation temperature
for vortices of extendedness $k~a$ at respective $\beta_G^{(k)}=
k~\beta_G^{(1)}$. In order to prevent bigger finite size effects, now we
kept the physical volume constant using $(k~L)^3$ lattices.  The
percolation probability $C$ and the densities are shown in Figure~4 {\it
vs.} $\beta_H$ as measured for $k=2$ on a $32^3$ lattice.  From all
critical $\beta_{Hc}^{(k)}$ values we estimate $T^{\mathrm{perc}}=170$ or
130 GeV (without or with $t$-quarks). This corresponds to a Higgs mass
$M_H=94$ or 103 GeV ({\it cf.} $M_H^*=100$ GeV). 

\begin{figure}[!htb]
\begin{minipage}{7.5cm}
\begin{center}
\epsfig{file=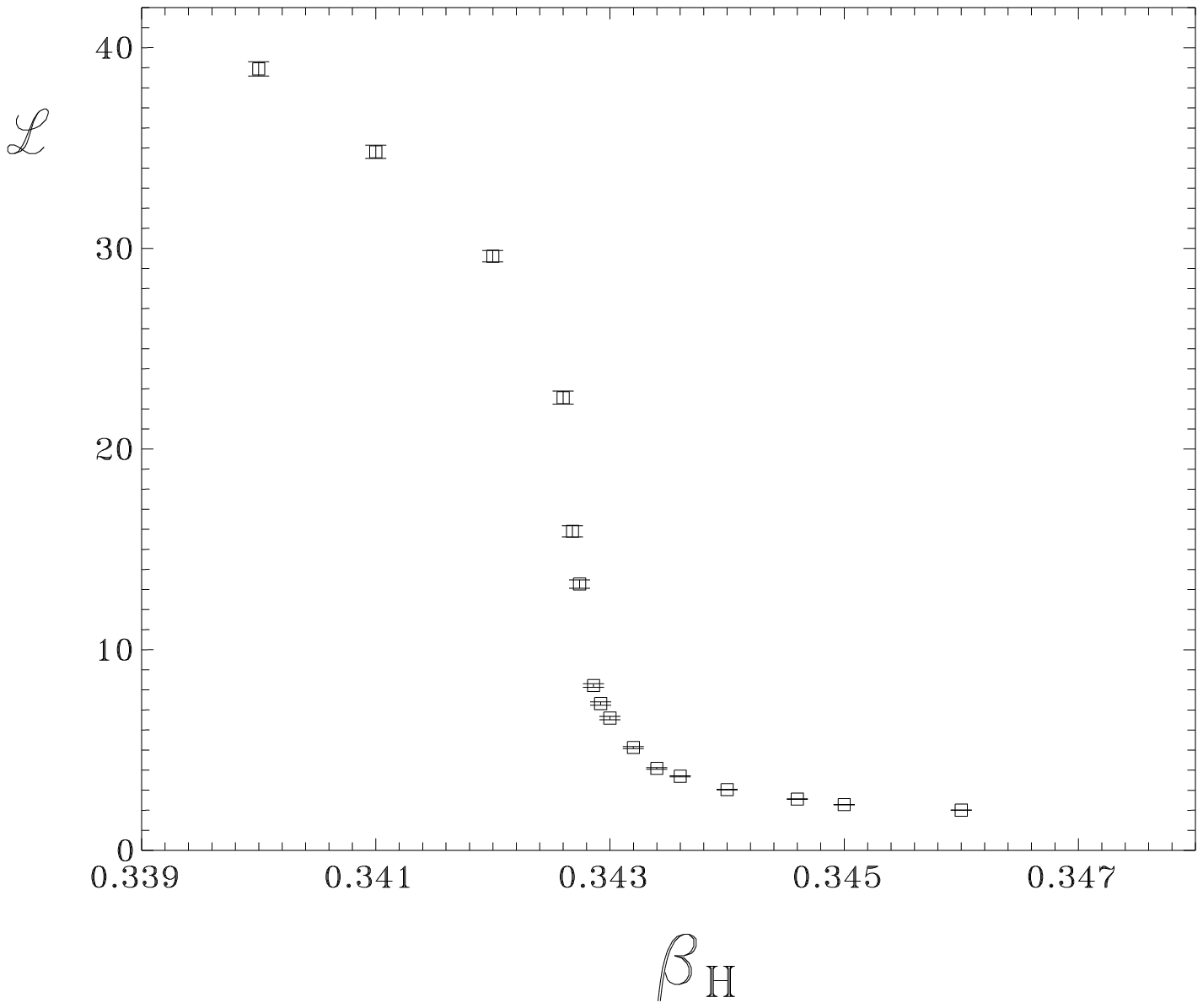,width=3.7cm,height=3.7cm,angle=0}
\epsfig{file=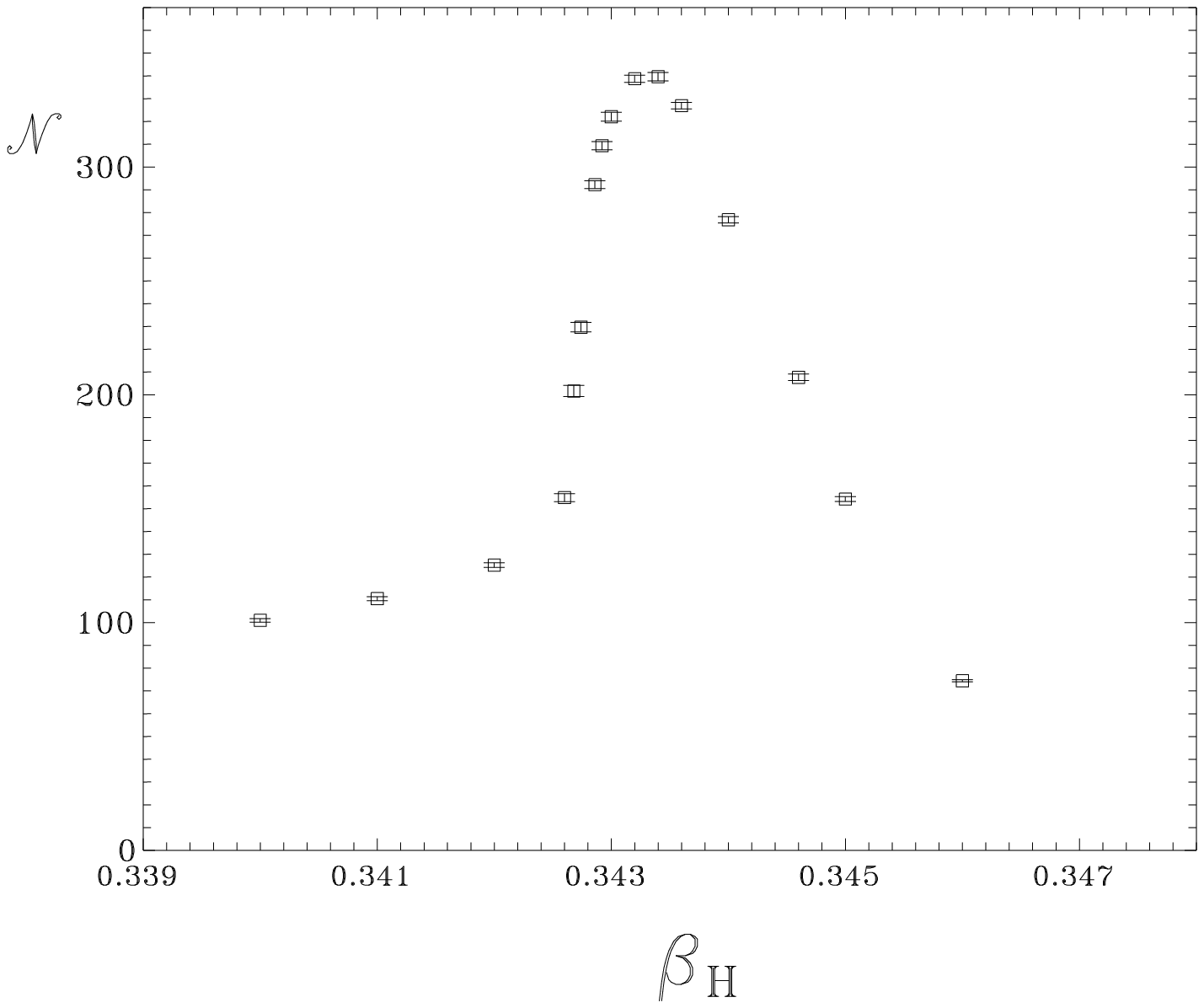,width=3.6cm,height=3.7cm,angle=0}
\vspace{-10mm}
\caption[]{\small
Average length $\cal L$ per $Z$--vortex cluster (left) and 
average number ${\cal N}$ of mutually disconnected $Z$--vortex clusters 
per lattice configuration $k=2$ objects on a
$32^3$ lattice at $\beta_G=16$.}
\end{center}
\end{minipage}
\end{figure}

The cluster structure of the configurations changes drastically at this
temperature. We illustrate this for $k=2$ vortices in Figure~5 with the
average number of clusters and the average length per cluster {\it vs.}
$\beta_H$. 

Our results obtained in an effective $3D$ equilibrium
approach suggest a scenario with 
a (few) percolating clusters decaying into smaller ones, vortex rings and
monopoliums, a process to be described by a kinetic approach to the crossover. 
Note that the average length per cluster has a long tail at a 
level of two. We propose to interpret the half density $\rho_v/2$ as the 
density of spalerons on the broken side of the crossover. We support this 
conjecture by what we have found~\cite{weSphal} to be the signature of a 
{\it classical} lattice sphaleron with respect to the new ($Z$-vortex and 
Nambu monopole) degrees of freedom.
Figure~6 shows one of the solutions of Ref. \cite{vanBaal} with a Nambu
monopole-antimonopole pair in the center, compared with a view in the 
maximally Abelian gauge. 

\begin{figure}[!htb]
\begin{minipage}{7.5cm}
\begin{center}
\begin{tabular}{c}
\epsfig{file=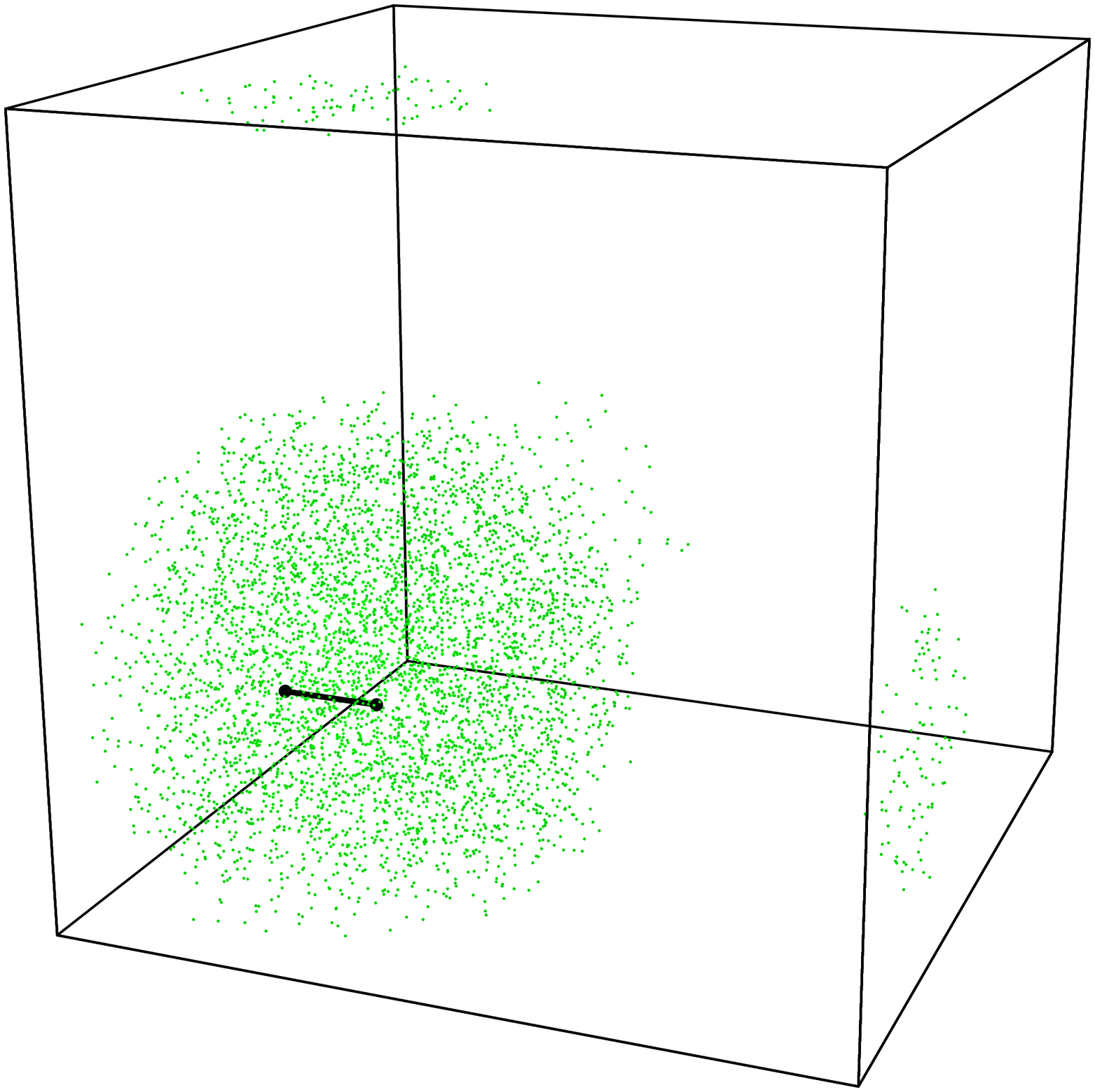,width=5.7cm,height=5.7cm,angle=0}\vspace{-8mm}\\
\epsfig{file=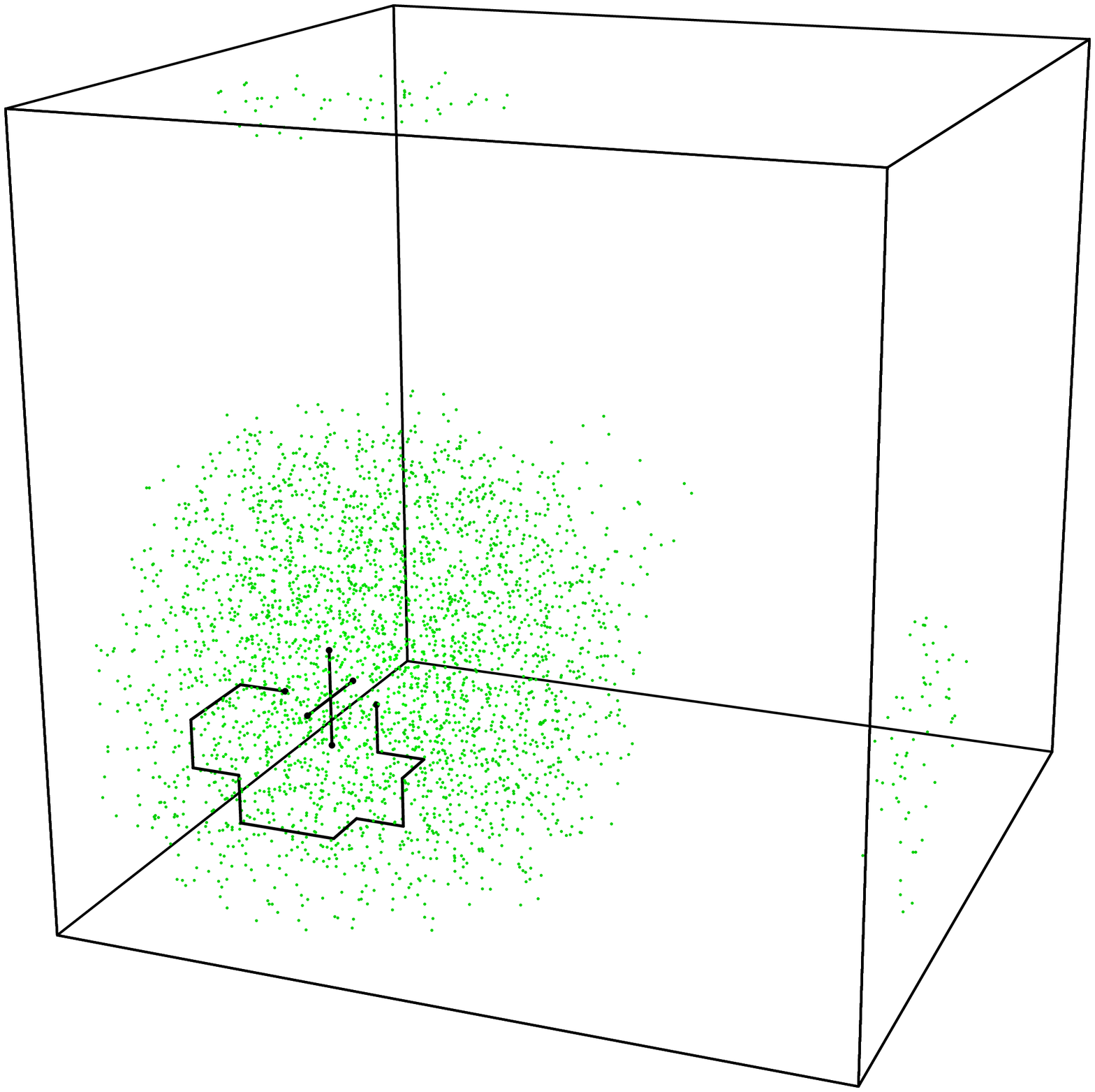,width=5.7cm,height=5.7cm,angle=0}\\
\end{tabular}
\vspace{-10mm}
\caption[]{\small
Classical lattice sphaleron with a Nambu monopole-anti-monopole pair
bound by a $Z$-vortex string in the center (above) and with Abelian monopoles
and Abelian vortices in the maximally Abelian gauge (below). 
The density of the clouds is
proportional to the decrease in Higgs field modulus.}
\end{center}
\end{minipage}
\end{figure}

One might argue that the defects counted by the lattice operators
(\ref{jN})  and (\ref{SigmaN}) might be mere artifacts having nothing in
common with the vortex solutions in the continuum. Figure~7 shows that
this is not the case. Here the average gauge field action per
pla\-quet\-te and the Higgs field modulus squared averaged over the
corners of a plaquette is shown as function of $\beta_H$, separately for
plaquettes with
$\sigma_p \ne 0$ and $\sigma_p=0$.
The data refer to $M_H=100$ GeV and $\beta_G=8$. 

One can clearly see that the modulus of the Higgs field is smaller near to
the vortex trajectory than outside the vortex for all values of the
coupling $\beta_H$. Moreover, the value of the gauge field energy near to
the vortex center is larger than in the bulk (the increasing error of the
gauge field action inside the vortex at the largest $\beta_H$
reflects the small number of plaquettes with non-vanishing vorticity).
\vspace{-6mm}
\begin{figure}[!htb]
\vspace{4mm}
\begin{minipage}{7.5cm}
\begin{center}
\epsfig{file=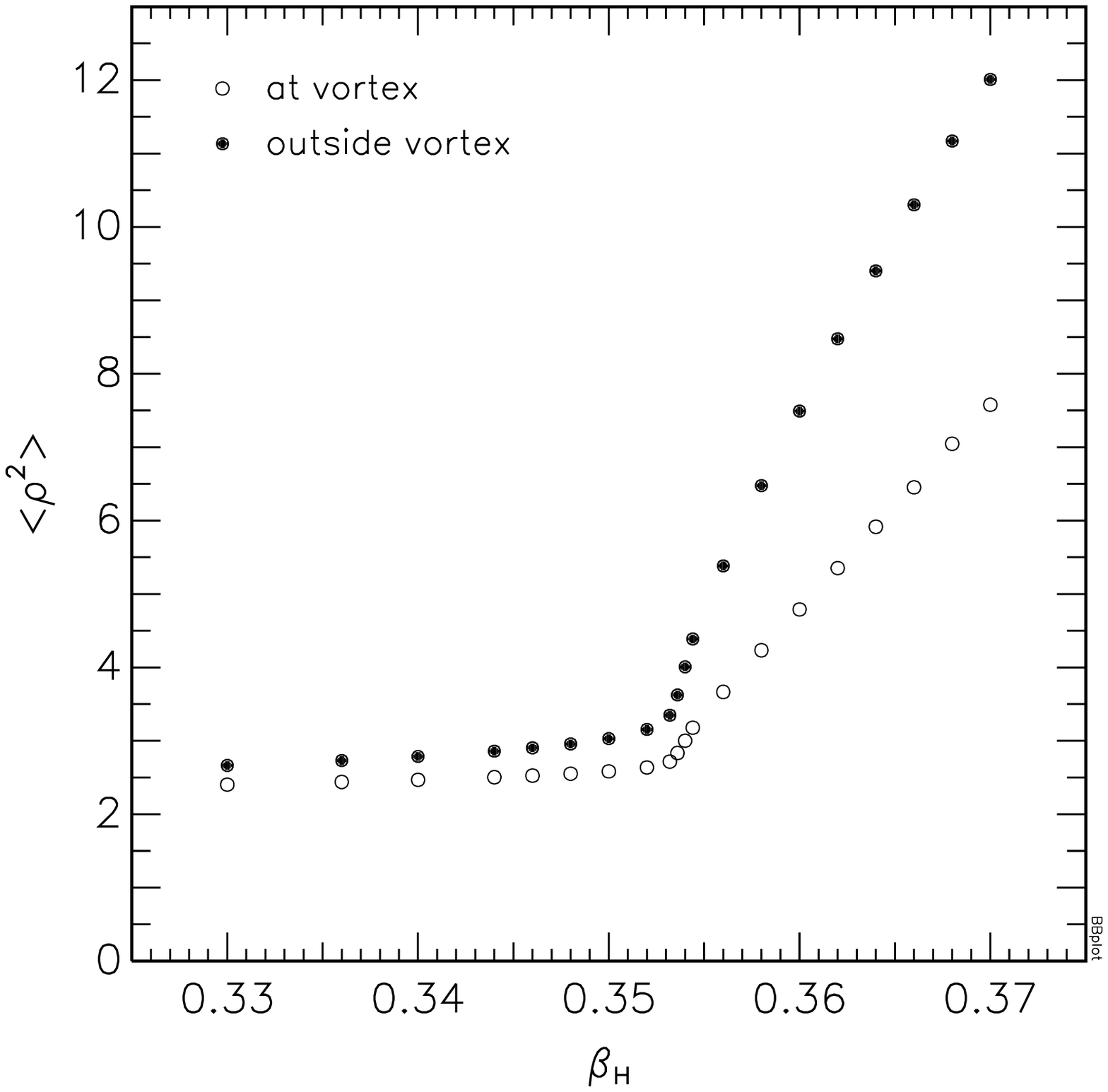,width=3.7cm,height=3.7cm,angle=0}
\epsfig{file=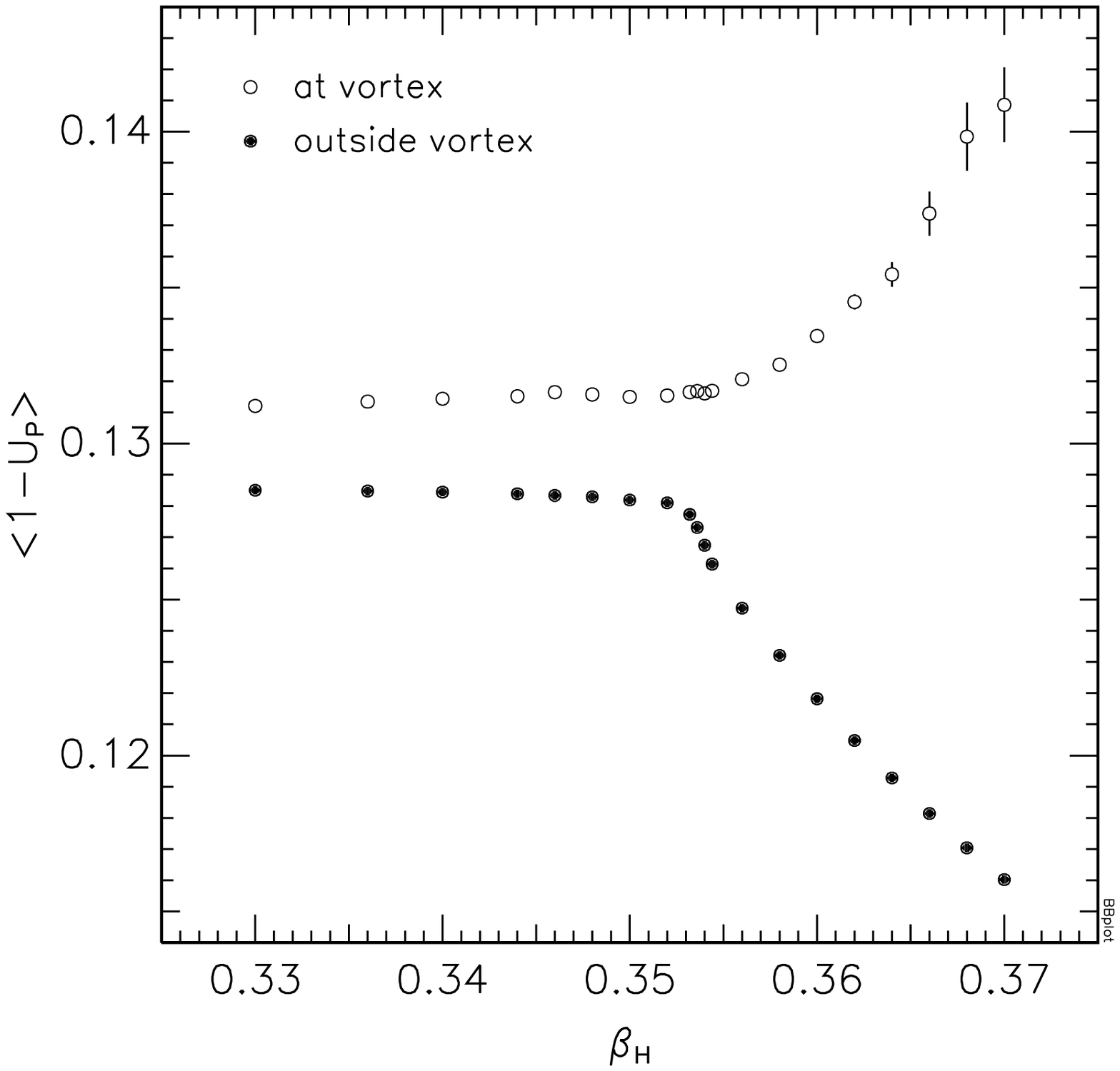,width=3.6cm,height=3.7cm,angle=0}
\vspace{-10mm}
\caption[]{\small
Average squared Higgs modulus (left) inside and outside
an elementary ($k=1$) $Z$--vortex  
on both sides of the
percolation transition. Right:
the same for gauge field energy.} 
\end{center}
\end{minipage}
\end{figure}

These results show that $Z$--vortices are physical objects which resemble
the characteristic features of the classical vortex solutions in
continuum: the vortices carry the excess of the gauge field energy and the
Higgs modulus decreases near to the vortex center.

\section{OUTLOOK}

In future we have to complete these studies by a more systematic
exploration of the continuum behavior/size distribution of defects.
Furthermore we plan to extend our considerations to more realistic models
with non-zero Weinberg angle. 
Last but not least, Euclidean simulations seem to be necessary to
clarify the connection of the dynamics of vortices with the evolution of
the Chern-Simons number. This may shed some light on the
baryogenesis scenario \cite{StringScenario} based on the dynamical
properties of the $Z$--vortex string network.

\section*{ACKNOWLEDGMENTS}

The authors have be\-ne\-fited from dis\-cus\-sions with  L.~McLerran,
M.~M\"ul\-ler\--Pre\-uss\-ker, M.~I.~Polikarpov, V.~A.~Rubakov,
K.~Rummukainen, K.~Selivanov, T.~Suzuki and P.~G.~Tinyakov. We are very
grateful to Pierre van Baal and Margarita Garcia--Perez for providing the
sphaleron lattice configurations. 

M.~N.~Ch. and F.~V.~G. were partially supported by the grants
INTAS-96-370, INTAS-RFBR-95-0681, RFBR-96-02-17230a and
RFBR-96-15-96740.

\end{document}